# Magnetization processes in non single domain magnetite particles

N. A. Usov, O. N. Serebryakova

*Pushkov Institute of Terrestrial Magnetism, Ionosphere and Radio Wave Propagation, Russian Academy of Sciences, IZMIRAN, 108480, Troitsk, Moscow, Russia*

**Abstract** Quasi-static hysteresis loops of spherical and spheroidal magnetite nanoparticles with semi-axes ratio $a/b$ = 1.5 and 2.0 with different types of combined magnetic anisotropy are calculated using numerical simulation. For particles of each type the critical diameters $D_{cr}$ are determined so, that above $D_{cr}$ the magnetization curling becomes the easiest mode of particle magnetization reversal. Hysteresis loops are calculated both for single-domain nanoparticles in the diameter range $D_{cr} < D \leq D_c$, and for vortex particles with diameters $D > D_c$, where $D_c$ is the single-domain diameter. The results obtained are compared with the usual hysteresis loops of particles with diameters $D \leq D_{cr}$. Hysteresis loops of dilute oriented assemblies are studied for various angles of the external magnetic field relative to the particle symmetry axis. For the corresponding randomly oriented assemblies the hysteresis loops are obtained by averaging over this angle. It is shown that the magnetization reversal of nanoparticles studied occurs through the nucleation of the curling mode, which can be accompanied by the formation of various vortices in finite intervals of the external magnetic field. The remanent magnetization and coercive force of oriented and non-oriented dilute assemblies of magnetite nanoparticles with different aspect ratios are determined as the functions of the transverse particle diameter.

## 1. Introduction

Assemblies of magnetic iron oxides nanoparticles with diameters on the order of or larger than single domain diameter have recently attracted interest due to various applications in biomedicine [1–5]. The properties of inhomogeneous micromagnetic states realized in magnetite particles of submicron sizes are also of great importance for paleomagnetic studies [6-14]. It is well known [6] that magnetite nanoparticles of submicron size are often found in dispersed natural assemblies formed in volcanic rocks. It was suggested [11,12] that nearly uniformly magnetized vortex cores in submicron magnetite particles can make a significant contribution to the remanent magnetization of rocks. Furthermore, since the directions of vortex cores are separated by high potential barriers, the remanent magnetization of an assembly of vortices could be stable over billions of years [13]. Note that vortex type distributions in non-single-domain magnetite nanoparticles can be directly observed using electron holographic methods [7,11,14].

Recently [15], various stable micromagnetic states existing in zero external magnetic field in spheroidal magnetite nanoparticles with a semi-axes ratio $a/b$ = 0.5 – 2.0 have been calculated using the Landau – Lifshitz – Gilbert (LLG) equation in the range of transverse particle diameters $D = 2b \leq 120$ nm. In this work we calculate quasi-static hysteresis loops of dilute assemblies of such particles with the aspect ratios $a/b$ = 1.0, 1,5 and 2.0, respectively. As in paper [15], two characteristic types of orientation of cubic anisotropy axes of magnetite with respect to the particle symmetry axis are considered. In the first case, the particle symmetry axis is parallel to one of the hard axes of cubic anisotropy of magnetite. In the second case, it is parallel to one of the easy anisotropy axes.

For both types of combined magnetic anisotropy the hysteresis loops of dilute oriented assemblies of magnetite nanoparticles are constructed depending on the particle transverse diameter for different directions of the external magnetic field relative to the particle symmetry axis. Hysteresis loops of dilute, randomly oriented assemblies of nanoparticles are then obtained by averaging the loops of oriented assemblies along the external magnetic field directions. The results of the calculations are compared with well-known hysteresis loops of single-domain nanoparticles with coherent rotation of magnetization [16-19]. For brevity, these loops will be called as the hysteresis loops of Stoner–Wohlfarth (SW) type.

The calculations performed show that the magnetization reversal of spheroidal magnetite nanoparticles with diameters greater than the single-domain one significantly depend on the direction of the applied magnetic field relative to the particle symmetry axis, the existence of metastable micromagnetic states in such particles in low magnetic fields being also important. Randomly oriented assemblies of vortex magnetite particles have a reduced coercive force, which decreases with increasing transverse particle diameters. At the same time, in the range of diameters studied, the remanent magnetization of such assemblies is close to that for the corresponding SW assemblies with coherent magnetization rotation.

**2. Numerical simulation**

Quasi-static hysteresis loops of magnetite nanoparticles are calculated in this work by solving the dynamic LLG equation with phenomenological damping [19-21]. The particle saturation magnetization is taken to be $M_s$ = 450 emu/cm$^3$, cubic magnetic anisotropy constant $K_c$ = -10$^5$ erg/cm$^3$, and exchange constant $C = 2A = 2\times10^{-6}$ erg/cm. The magnetic damping parameter is given by $\kappa$ = 0.5. Spheroidal particles are assumed to be elongated along the $Z$ axis of the Cartesian coordinates, with $a$ and $b$ being the longitudinal and transverse semi-axis of the particle, respectively, $D = 2b$ is the transverse particle diameter. Calculations of quasi-static hysteresis loops are carried out in the range of transverse particle diameters from 60 to 110 nm.

The cubic magnetic anisotropy energy density of a magnetite particle has the form [15]

$$w_a = -|K_c|\left((\vec{\alpha}\vec{e}_1)^2(\vec{\alpha}\vec{e}_2)^2 + (\vec{\alpha}\vec{e}_1)^2(\vec{\alpha}\vec{e}_3)^2 + (\vec{\alpha}\vec{e}_2)^2(\vec{\alpha}\vec{e}_3)^2\right), \qquad (1)$$

where $\vec{\alpha}$ is the unit magnetization vector, and $(\vec{e}_1, \vec{e}_2, \vec{e}_3)$ is the orthogonal set of unit vectors that determines the orientation of the cubic easy anisotropy axes of the magnetite nanoparticle. For the case of combined anisotropy of 'd' type, the vectors $(\vec{e}_1, \vec{e}_2, \vec{e}_3)$ are oriented parallel to the Cartesian coordinate axes. In this case the easy axes of cubic anisotropy of magnetite are directed along the unit vectors $(\pm 1/\sqrt{3}, \pm 1/\sqrt{3}, \pm 1/\sqrt{3})$, the symmetry axis of the particle being parallel to the hard axis of cubic anisotropy. On the other hand, for the case of combined anisotropy of 'z' type, the orthogonal set $(\vec{e}_1, \vec{e}_2, \vec{e}_3)$ is oriented so [15] that one of the easy axes of cubic anisotropy is parallel to the particle symmetry axis.

For numerical simulation a spheroidal nanoparticle is approximated by a set of $N \sim 10^4$ small cubic elements with an edge size of 3 - 5 nm, small compared to the exchange length of magnetite, $L_{ex} = \sqrt{C}/M_s$ = 31.4 nm. The calculation of the quasi-static hysteresis loop begins in an external magnetic field $H$ = 1500 - 2000 Oe, sufficient for complete magnetic saturation of a spheroidal magnetite nanoparticle. The direction of the external magnetic field is specified by the spherical angles $\omega$ and $\psi$. The dynamic evolution of the initial magnetization distribution in a particle is traced to the final state, which is assumed to be stable in a given magnetic field under the condition

$$\max_{(1\leq i \leq N)} \left\|\vec{\alpha}_i \times \vec{H}_{ef,i}/\|\vec{H}_{ef,i}\|\right\| < 10^{-6}. \qquad (2)$$

Here $\vec{\alpha}_i$ and $\vec{H}_{ef,i}$ are the unit magnetization vector and effective magnetic field in the $i$-th numerical cell of the particle, respectively. After reaching a stationary state, the amplitude of the external magnetic field decreases by a small amount $\Delta H$ = 1 - 2 Oe and a new stationary state is calculated in a similar way.

**3. Quasi-static hysteresis loops**

**3.1. Spheroids with anisotropy of 'd' type**

*a/b* = 1.5

The quasi-static hysteresis loops of particles with 'd' type anisotropy and aspect ratio $a/b$ = 1.5, calculated numerically, are shown in Fig. 1 for different angles $\omega$ of magnetic field relative to the particle symmetry axis. The dependence of the hysteresis loops on the azimuthal angle $\psi$ is insignificant. Various dots in Fig. 1 show the descending branches of the hysteresis loops for particles with transverse diameters $D$ = 62, 74 and 82 nm, respectively. For comparison, in Fig. 1 solid curves show the hysteresis loops of SW type, which do not depend on the particle diameter [16-19].

It is known [17, 18] that the magnetization reversal of a single-domain particle is carried out by uniform rotation in the domain $D \le D_{cr}$. For the case $\omega = 0$ the critical diameter $D_{cr}$ for spheroidal particles can be determined analytically [18,22,23]. The critical field for magnetization reversal of a particle by uniform rotation for the case $\omega = 0$ is [16-18]

$$H_{cr}(0) = -\frac{2K_{ef}}{M_s}. \qquad (3)$$

Here

$$K_{ef} = M_s^2(\pi - 0.75 N_z) - |K_c|, \qquad (4)$$

is the effective anisotropy constant of an elongated spheroid with combined anisotropy of 'd' type [15], $N_z = N_z(a/b)$ being the longitudinal demagnetizing factor of the spheroid. On the other hand, the nucleation field of the curling mode for the case under consideration is equal to [15]

$$H_{curl}(0) = \frac{2|K_c|}{M_s} - C\frac{q_{11}^2(a/b)}{b^2 M_s} + N_z M_s. \qquad (5)$$

The numerical function $q_{11}(a/b)$ in Eq. (5) is proportional to the minimum root of the derivative of the radial spheroidal function $R_{11}(c_{11}, \xi)$ on the spheroid surface [24]. Equating formulas (3) and (5), one obtains the critical transverse diameter for elongated spheroid at $\omega = 0$ as follows

$$D_{cr} = \frac{2 q_{11}(a/b)}{\sqrt{N_z(1 + 2(|K_c| + K_{ef})/N_z M_s^2)}} L_{ex}. \qquad (6)$$

For magnetite nanoparticles with $a/b = 1.5$ Eq. (6) gives the value $D_{cr} = 58.7$ nm, whereas the transverse single-domain diameter determined [15] for particles of this type is $D_c = 66.0$ nm. Thus, a single-domain particle with $D = 62$ nm falls into the diameter range $D > D_{cr}$, where the curling mode is the easiest one. However, particles with transverse diameters $D = 74$ and 82 nm are in the vortex_z state in zero magnetic field [15]. Note, the axis of the vortex_z is parallel to the particle symmetry axis.

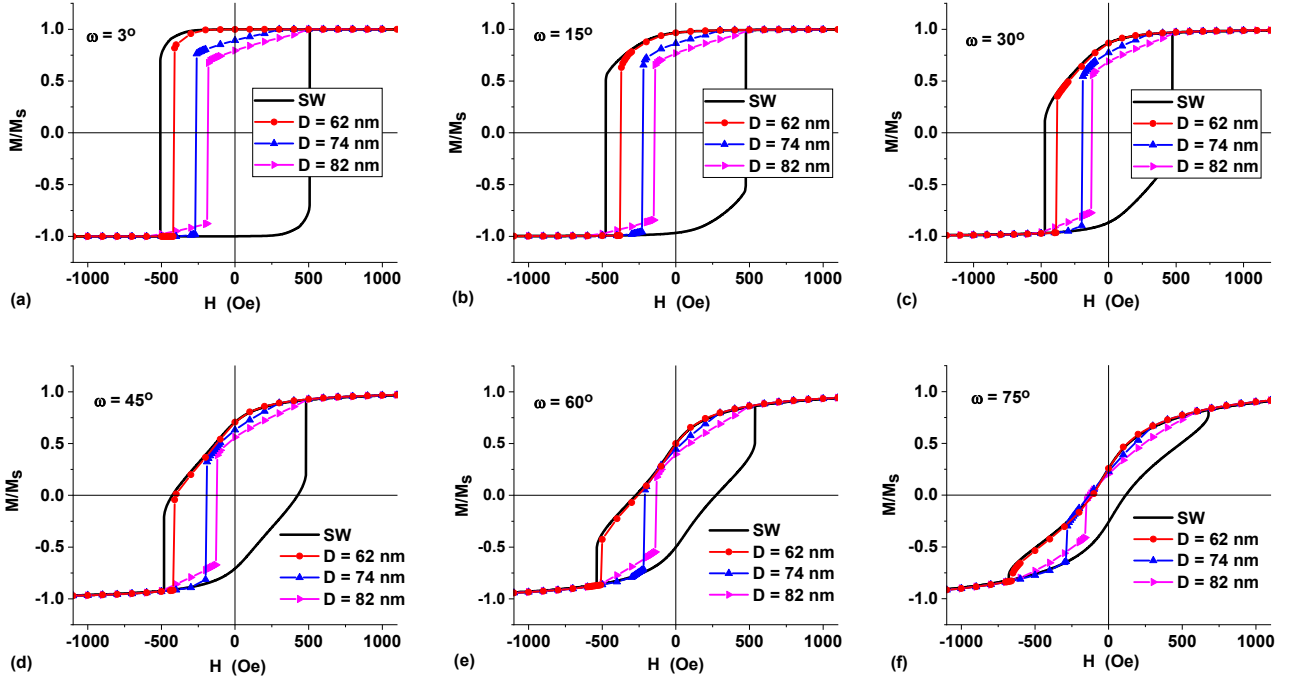

Fig. 1. Descending branches of quasi-static hysteresis loops of particles with 'd' type anisotropy, aspect ratio $a/b = 1.5$ and diameters $D = 62$, 74 and 82 nm (different dots) for angles $\omega = 3°$, 15°, 30°, 45°, 60° and 75° in comparison with SW hysteresis loops (solid curves).

Fig. 1 shows that when the external magnetic field decreases from a large positive value to a negative one, the hysteresis loops for the particle with $D$ = 62 nm for all angles $\omega$ coincide with the SW loops up to the critical nucleation field of the curling mode. However, as Figs. 1a – 1d show, due to the earlier nucleation of the curling mode, the switching fields $H_{sw}$ of this particle decreases compared to that for the SW hysteresis loops. Nevertheless, according to Figs. 1e, 1f, for angles $\omega \geq 60°$ the difference in the switching fields of these particles becomes insignificant.

At the same time, as can be seen from Fig. 1a, for vortex particles with diameters $D$ = 74 and 82 nm the nucleation of the vortex at $\omega$ = 3° occurs already in positive magnetic fields $H_{vn}$ = 300 Oe and 500 Oe, respectively. As a result of the vortex formation the remanent magnetization for these particles at $\omega \leq 30°$ noticeably decreases and their hysteresis loops deviate significantly from the SW loops. As shown in Figs. 1 and 2a, the switching fields $H_{sw}$ for vortex nanoparticles decrease with increasing particle diameter and turn out to be considerably less than those for the particle with $D$ = 62 nm. A significant decrease in switching fields can be considered as a characteristic property of vortex particles.

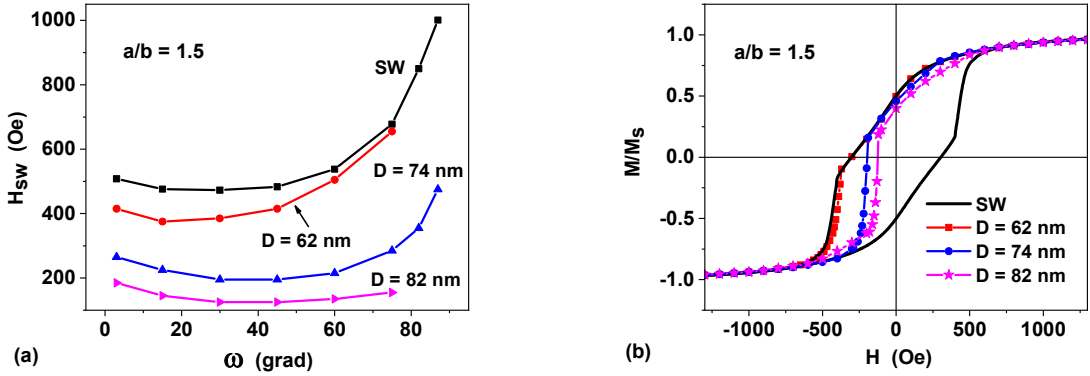

Fig. 2. a) Angular dependence of switching fields of spheroidal magnetite nanoparticles with 'd' type anisotropy and aspect ratio $a/b$ = 1.5 depending on the transverse particle diameter. b) Hysteresis loops of dilute randomly oriented assemblies of spheroidal particles with the same anisotropy for particles of different diameters.

In Fig. 2b various dots show the hysteresis loops of randomly oriented, dilute assemblies of magnetite particles with transverse diameters $D$ = 62, 74 and 82 nm, in comparison with the corresponding SW loop, shown as a solid curve. One can see that the hysteresis loop for particles with $D$ = 62 nm differs only slightly from the SW loop, while the hysteresis loops for vortex particles with diameters $D$ = 74 and 82 nm have a significantly lower coercive force. Nevertheless, the remanent magnetization of randomly oriented assemblies with aspect ratio $a/b$ = 1.5 differs only slightly from that for the SW loop. This is mainly due to the fact that in the range of particle diameters studied, the decrease in the total particle magnetization during the vortex formation is small.

Fig. 3 shows the magnetization reversal process of vortex particle with diameter $D$ = 82 nm for two typical directions of the magnetic field relative to the particle symmetry axis. In this and subsequent graphs of this type the external magnetic field is oriented in the XZ plane, so that the azimuthal angle $\psi$ = 0. Curves 1 in Fig. 3 show the projection of the reduced particle magnetization $M$ onto the external magnetic field direction, whereas curves 2 show the total reduced particle magnetization $M_t = \sqrt{M_x^2 + M_y^2 + M_z^2}$ as a function of external magnetic field. The inserts in Figs. 3a and 3b show the evolution of the reduced magnetization components $M_x$, $M_y$ and $M_z$ averaged over the particle volume, when the magnetic field decreases from $H$ = 1500 Oe to – 1500 Oe.

Curve 2 in Fig. 3a shows that at $\omega$ = 45° the vortex nucleation in the particle occurs at $H_{nv}$ = 500 Oe, since for $H < H_{nv}$ the total particle magnetization $M_t$ becomes less than 1. Insert in Fig. 3a shows that after the vortex appearance, in the field range from 500 Oe to zero, the $M_z$ magnetization component changes relatively weakly, while the $M_x$ component drops to zero.

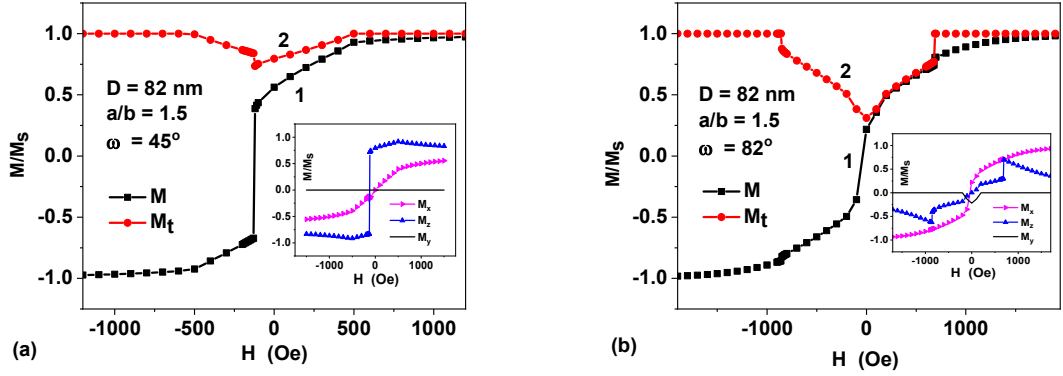

Fig. 3. Magnetization reversal of vortex particle with diameter $D = 82$ nm for magnetic field directions $\omega = 45°$ (a) and $\omega = 82°$ (b), respectively. Curves 1 show the projection of the particle magnetization $M$ onto the external magnetic field direction, whereas curves 2 give the evolution of the total particle magnetization $M_t$. The insets show the dependence of the reduced magnetization components $M_x$, $M_y$ and $M_z$ on the applied magnetic field.

Thus, in this field region the vortex is compressed and its axis gradually rotates to Z axis. This happens because, according to the energy diagram shown in Fig. 1b of [15], the vortex_z is the lowest energy state of the given particle at $H = 0$. With a further decrease in the magnetic field, at $H = -125$ Oe, the vortex loses stability, the $M_z$ component abruptly jumps and the projection of the particle magnetization changes sign. Then, up to the field $H = -500$ Oe, the vortex is gradually displaced from the particle. Due to the symmetry of the magnetization distribution the $M_y$ component remains close to zero when the external field changes in the XZ plane. It is interesting to note that in the given process the decrease in the total particle magnetization $M_t$ due to the vortex formation does not exceed 27%.

    The magnetization reversal in this particle occurs in a similar way for angles $\omega \leq 75°$, but it changes for angles $82° \leq \omega \leq 90°$. As Fig. 3b shows, unlike Fig. 3a at $\omega = 82°$ the formation of a transverse vortex_p whose axis is perpendicular to Z axis occurs in the field range from 680 Oe to zero. Indeed, as inset in Fig. 3b shows, the $M_z$ magnetization component of the particle is close to zero while $M_y$ component is finite in the interval $|H| < 100$ Oe. Therefore, the vortex axis is oriented in the XY plane at a certain angle $\psi$ relative to the X axis. The total vortex magnetization $M_t$ drops to 31% at $H = 0$. With a further decrease in the field, the vortex magnetization returns to the XZ plane and $M_y$ component disappears. The displacement of the vortex from the particle occurs at the symmetric point $H = -680$ Oe, which is accompanied by a jump in the $M_z$ component.

    Thus, the magnetization reversal of the vortex particle with $D = 82$ nm for angles $\omega \leq 75°$ occurs through the formation of vortex_z in low magnetic fields. On the contrary, the formation of the vortex_p is observed in low fields for $\omega \geq 82°$. Note, the vortex_p is the metastable state of this particle at $H = 0$, [15]. For the vortex particle with $D = 74$ nm similar behavior occurs for angles $85° < \omega < 90°$.

*a/b = 2.0*

    For aspect ratio $a/b = 2.0$ the quasi-static hysteresis loops were studied for particles with transverse diameters $D = 64$, 72, 80, 84, and 90 nm. For magnetite nanoparticles with $a/b = 2.0$ from Eq. (6) one finds the value $D_{cr} = 55.8$ nm. The single-domain diameter for particles of this type has been determined [15] to be $D_c = 72.0$ nm. Thus, particles with diameters $D = 64$ and 72 nm are single-domain, but fall into the domain $D > D_{cr}$. However, the vortex_z is the ground state of particles with diameters $D = 80$, 84 and 90 nm at $H = 0$.

    Fig. 4 shows the descending branches of the hysteresis loops of nanoparticles of different transverse diameters for characteristic directions of the external magnetic field relative to the particle symmetry axis.

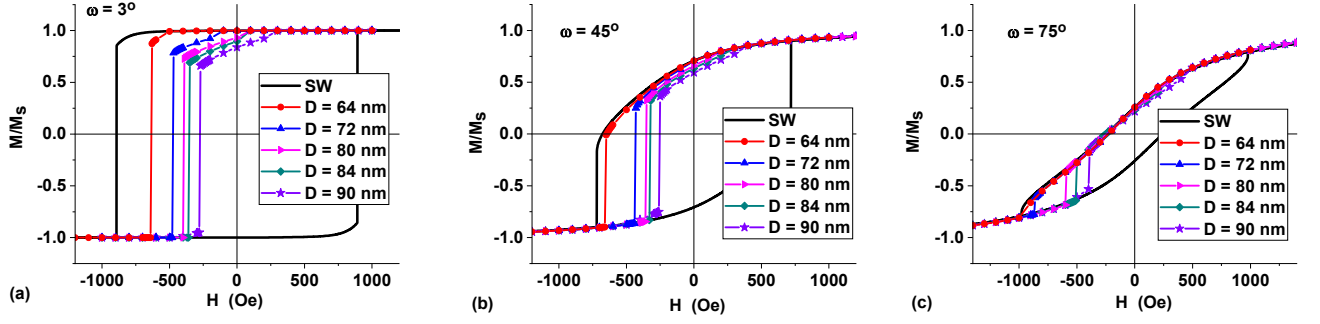

Fig. 4. Descending branches of hysteresis loops of nanoparticles with aspect ratio $a/b = 2.0$ and type 'd' anisotropy for different directions of the external magnetic field with respect to the particle symmetry axis: a) $\omega = 3°$, b) $\omega = 45°$ and c) $\omega = 75°$, respectively. SW hysteresis loops are shown as solid curves.

As can be seen in Fig. 4a, at $\omega = 3°$ single-domain nanoparticles with $D = 64$ and 72 nm remain uniformly magnetized when the magnetic field decreases to zero. With a further decrease in the field, a vortex state develops in these particles in magnetic fields $H_{nv} = -500$ Oe and $-100$ Oe, respectively. As a result, the projection of the reduced particle magnetization onto the field direction decreases. However, the abrupt magnetization reversal of these particles occurs only at $H_{sw} = -640$ Oe and $-480$ Oe, respectively. Thus, for these particles the appearance of the vortex state at small angles $\omega$ is not accompanied by immediate magnetization reversal.

On the other hand, for vortex nanoparticles with diameters $D = 80$, 84 and 90 nm at $\omega = 3°$ the vortex appears already in positive fields $H_{nv} = 100$, 120 and 300 Oe, respectively. The switching of the particle magnetization occurs in negative fields $H_{sw} = -400$, $-360$ and $-280$ Oe, respectively. It is interesting to note that, as Figs. 4b, 4c show, at angles $\omega \geq 45°$ the projection of magnetization onto the field direction, both for single-domain and vortex particles, remains close to the SW hysteresis loop up to the switching fields of these particles.

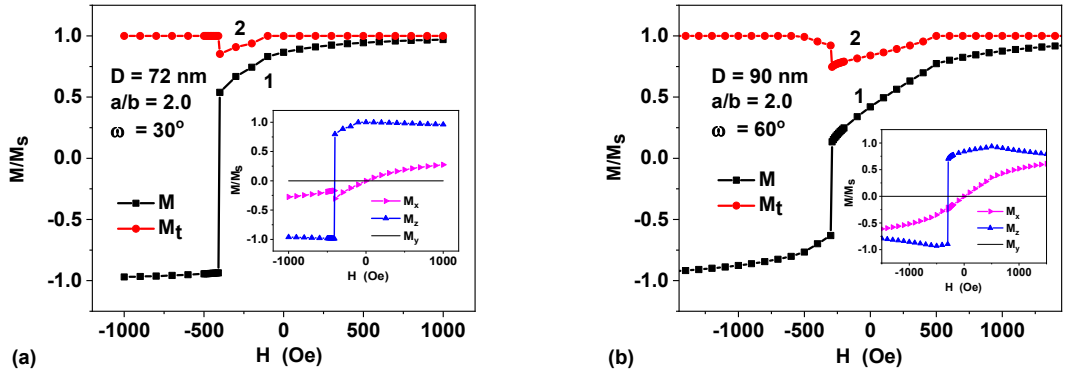

Fig. 5. The magnetization reversal of particles with aspect ratio $a/b = 2.0$, anisotropy type 'd' and transverse diameters $D = 72$ and 90 nm at the angles $\omega = 30°$ (a) and $\omega = 60°$ (b), respectively.

Details of the magnetization reversal of particles with aspect ratio $a/b = 2.0$ are shown in Fig. 5a for a single-domain particle with diameter $D = 72$ nm, and in Fig. 5b for a vortex particle with $D = 90$ nm, respectively. Curves 1 in Fig. 5 show the projection $M$ of the particle magnetization on the magnetic field direction, whereas curves 2 give the evolution of the total magnetization of the particle $M_t$. It is obvious that the magnetization reversal process of particles with diameters $D = 72$ and 90 nm differs only in the sense that for the single-domain particle the vortex arises in a negative field, $H_{nv} = -100$ Oe, but for vortex particle it appears in a positive field, $H_{nv} = 400$ Oe. In addition, the single-domain particle becomes uniformly magnetized after a jump of its $M_z$ component at $H_{sw} = -410$ Oe. In the vortex particle after a magnetization jump in the field $H_{sw} = -300$ Oe, a vortex with a negative core magnetization appears. It is slowly forced out of the particle and completely disappears only at $H = -500$ Oe.

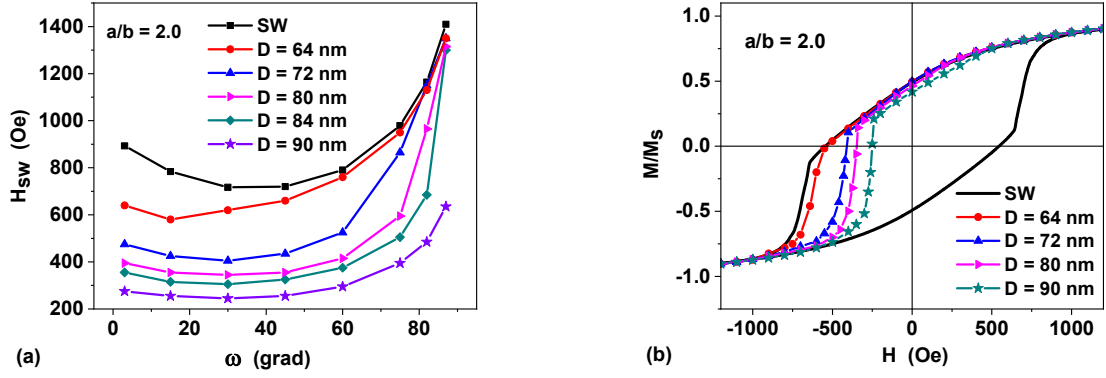

Fig. 6. a) Angular dependence of switching fields of spheroidal magnetite nanoparticles with 'd' type anisotropy and aspect ratio $a/b = 2.0$ for particles of different transverse diameters. b) Hysteresis loops of randomly oriented dilute assemblies of the same nanoparticles depending on the transverse particles diameter.

The magnetization reversal of vortex particles with aspect ratio $a/b = 2.0$ occurs through the formation of longitudinal vortex_z even for large angles $\omega \geq 75°$. It is the consequence of the fact that, as Fig. 1c of [15] shows, the transverse vortex_p for particles of this type is absent in the diameter range studied.

    Similar to Fig. 2a, Fig. 6a demonstrates a significant decrease in the switching fields of spheroidal particle with $a/b = 2.0$ with increasing transverse diameter. Fig. 6b shows that in the studied range of diameters even the hysteresis loops of vortex particles are quite close to the hysteresis loop of the randomly oriented SW assembly up to the switching fields. As a result, the remanent magnetizations of the assemblies studied turn out to be quite close to each other. However, the coercive force of assemblies of vortex particles quickly decreases with increasing transverse particle diameter.

### 3.2. Spheroids with 'z' type anisotropy

    For combined anisotropy of 'z' type the critical field for uniform magnetization rotation is also given by Eq. (3) with the difference that the effective anisotropy constant of elongated spheroid is now given by [15]

$$K_{ef} = M_s^2(\pi - 0.75 N_z) + 2|K_c|/3. \quad (7)$$

The nucleation field of the curling mode in this case is equal to

$$H_{curl}(0) = -\frac{4|K_c|}{3M_s} - C\frac{q_{11}^2(a/b)}{b^2 M_s} + N_z M_s. \quad (8)$$

Therefore, the critical diameter for the curling mode to be the easiest one at $\omega = 0$ is

$$D_{cr} = \frac{2 q_{11}(a/b)}{\sqrt{N_z\left(1 - 4|K_c|/3N_z M_s^2 + 2K_{ef}/N_z M_s^2\right)}} L_{ex}. \quad (9)$$

Taking into account Eqs. (4) and (7), it is easy to verify that Eqs. (6) and (9) actually coincide. Therefore, the previously obtained values $D_{cr} = 58.7$ nm and 55.8 nm for spheroids with $a/b = 1.5$ and $a/b = 2.0$, respectively, turn out to be valid also for particles with 'z' type anisotropy.

#### $a/b = 1.5$

    Calculations of hysteresis loops for the case of $a/b = 1.5$ and 'z' type anisotropy were carried out for particles with transverse diameters $D = 70$, 82 and 90 nm. For particles of this type the single-domain diameter was determined as $D_c = 86.6$ nm [15]. As the upper diameter for uniform rotation is $D_{cr} = 58.7$ nm, the particles with diameters $D = 70$ and 82 nm are single-domain, but fall in the range $D > D_{cr}$.

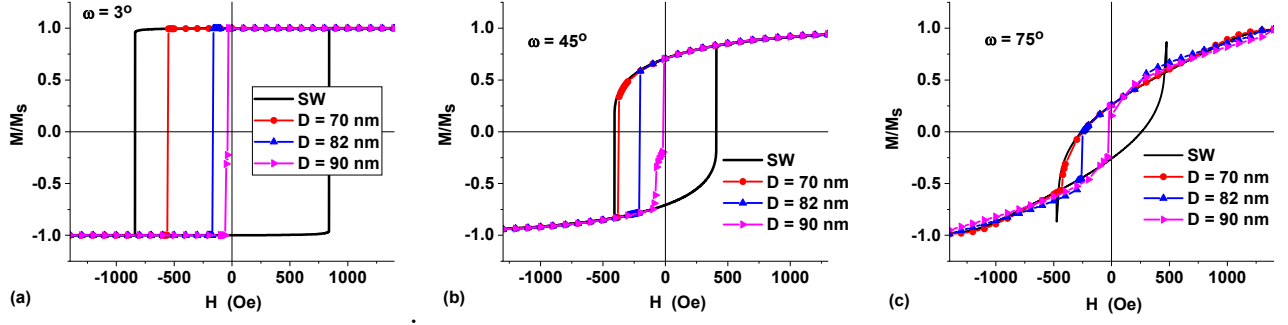

Fig. 7. Comparison of descending branches of hysteresis loops of particles with aspect ratio $a/b = 1.5$, 'z' type anisotropy and diameters $D = 70$, 82 and 90 nm for field angles $\omega = 3°$ (a) $\omega = 45°$ (b) and $\omega = 75°$ (c), respectively. SW hysteresis loops are shown as solid curves.

However, the lowest energy state of vortex particle with $D = 90$ nm in zero magnetic field is vortex_p, with the vortex axis perpendicular to $Z$ axis [15]. In addition, in particles of this type in the diameter range $D_c < D \leq 90$ nm there is also a metastable uniform state magnetized along $Z$ axis [15]. As we shell see, both of these states are involved in the magnetization reversal processes of particles of this type.

As Fig. 7 shows, as the field decreases from a large positive value, the hysteresis loops of single-domain particle with $D = 70$ nm at different angles $\omega$ coincide with SW hysteresis loops up to the critical field $H_{curl}$, when the nucleation of the curling mode occurs. Moreover, for $\omega \leq 30°$ the fields $H_{curl}$ are significantly less than the critical fields $H_{cr}$ for SW hysteresis loops. But for the angles $\omega \geq 45°$ the difference in the switching fields of these particles is small.

A single-domain particle with $D = 82$ nm for angles $\omega \leq 60°$ also undergoes abrupt magnetization reversal through the curling mode nucleation. An example of this behavior is shown in Fig. 8a for the case $\omega = 60°$. As curve 2 in Fig. 8a shows, this particle is uniformly magnetized both before and after the magnetization jump at $H_{sw} = -240$ Oe, since its total magnetization is unchanged, $M_t = 1$, in the entire field range from 1600 to $-1600$ Oe. However, for the angles $\omega \geq 75°$ the magnetization reversal process occurs in this single-domain particle with the participation of a vortex. Indeed, as Fig. 8b shows, at $\omega = 87°$ the magnetization of this particle is uniform and parallel to applied magnetic field in $H \geq 1000$ Oe. But in the field interval $100 \text{ Oe} \leq H \leq 1000$ Oe the total particle magnetization $M_t < 1$ since a vortex exists in this particle. Further, as curve 2 in Fig. 8b shows, in the field range $-320 \text{ Oe} \leq H \leq 100$ Oe the uniform magnetization with $M_z \approx 1$ is restored in this particle. Note that the uniform magnetization is the ground state of this particle at $H = 0$ [15]. Finally, at $H = -330$ Oe, the vortex with the opposite core magnetization appears in this particle again. With further decrease in the field, the axis of the vortex turns perpendicular to the particle symmetry axis and the vortex disappears at $H = -1100$ Oe.

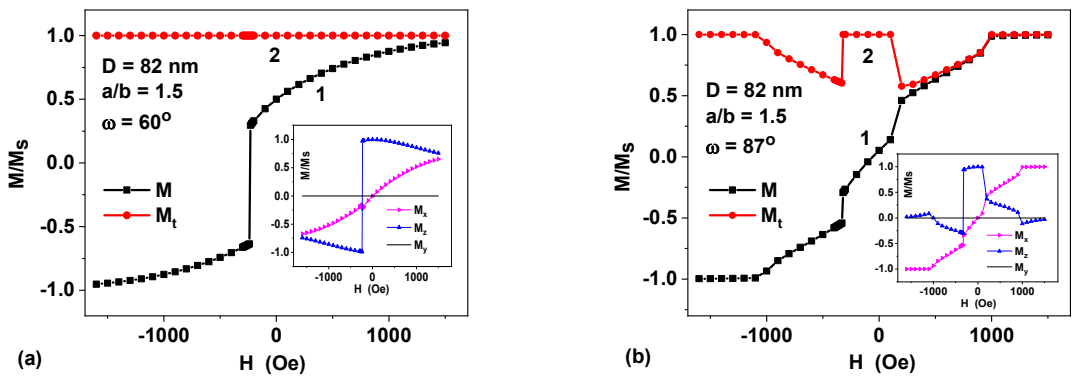

Fig. 8. The magnetization reversal process of a single-domain particle with 'z' type anisotropy, $a/b = 1.5$ and diameter $D = 82$ nm at $\omega = 60°$ (a) and $\omega = 87°$ (b), respectively.

Interestingly, a vortex particle with diameter $D$ = 90 nm for angles $\omega \leq 60°$ remains uniformly magnetized, except of relatively small intervals of negative fields, where transverse vortex_p exists in this particle. Note, vortex_p is the ground state for this particle at $H$ = 0 [15]. As an example, curve 2 in Fig. 9a shows that at $\omega$ = 30° a vortex_p exists in this particle in a small range of negative fields – 65 Oe ≤ $H$ ≤ - 5 Oe, where the total particle magnetization decreases to $M_t \approx 0.27$. Outside this interval, and in particular, at $H$ = 0, the particle remains uniformly magnetized. Thus, the magnetization reversal process occurs in this particle through uniform metastable state at $H$ = 0.

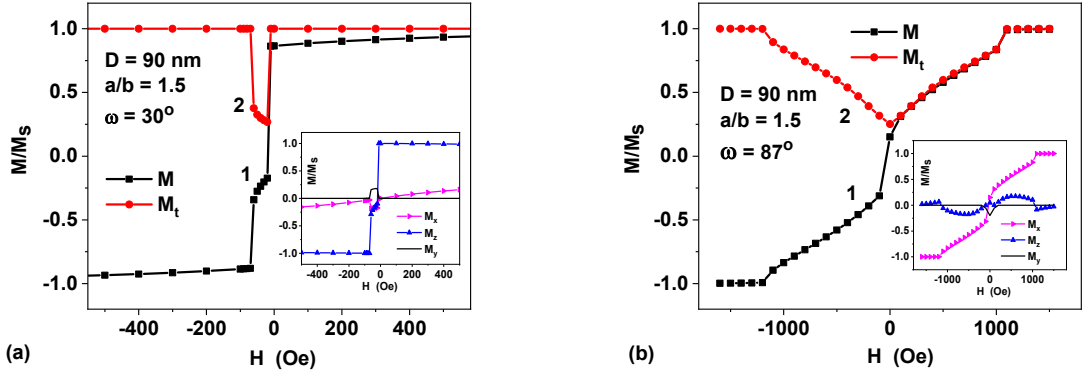

Fig. 9. The magnetization reversal process of a vortex particle with 'z' type anisotropy, $a/b$ = 1.5 and diameter $D$ = 90 nm at the angles $\omega$ = 30° (a), and $\omega$ = 87° (b), respectively.

On the other hand, the behavior of the particle changes significantly for angles $\omega \geq 75°$. As Fig. 9b shows, at $\omega$ = 87° and $H$ > 1100 Oe this particle is uniformly magnetized along applied magnetic field. The nucleation of the vortex occurs in this particle at $H_{nv}$ = 1100 Oe. In the interval from $H$ = 1000 Oe to zero, the vortex contracts and rotates in the *XZ* plane, so that in zero field the total particle magnetization decreases up to $M_t$ = 0.25. In a small field range $|H| \leq 100$ Oe the $M_x$ magnetization component changes sign, and with a further increase in the field in the negative direction, the vortex turns and unwinds. It completely disappears in the particle at $H$ = - 1200 Oe.

Thus, for angles $\omega \leq 60°$ the magnetization reversal of the vortex particle with diameter $D$ = 90 nm occurs through metastable uniform magnetization in zero field, and the transverse vortex_p exists only in a relatively small range of negative magnetic fields. On the other hand, for $\omega \geq 75°$ the vortex exists in the given particle in a wide field range, - 1200 Oe < $H$ < 1100 Oe.

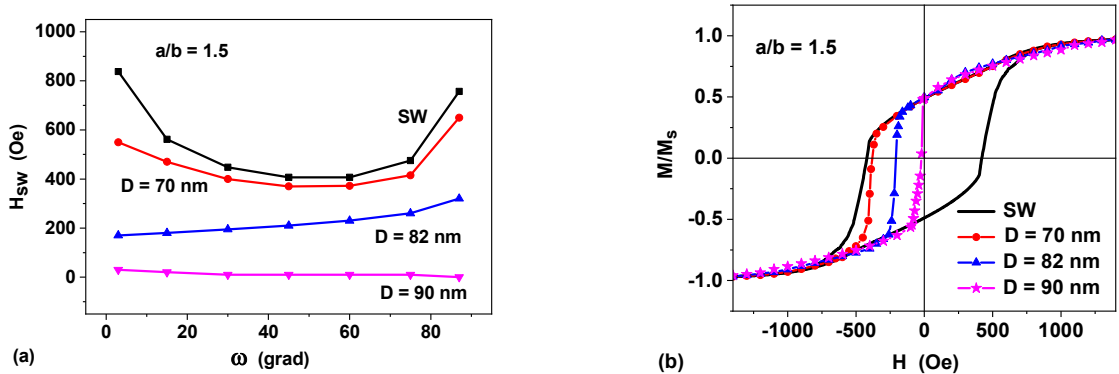

Fig. 10. (a) Angular dependence of switching fields of spheroidal magnetite nanoparticles with 'z' type anisotropy, $a/b$ = 1.5 and transverse diameters $D$ = 70, 82 and 90 nm, respectively. (b) Hysteresis loops of dilute randomly oriented assemblies of the same nanoparticles in comparison with the SW hysteresis loop.

Fig. 10a shows the angular dependence of the switching fields of spheroidal magnetite nanoparticles with 'z' type anisotropy, aspect ratio $a/b$ = 1.5 and diameters D = 70, 82 and 90 nm, whereas Fig. 10b shows the hysteresis loops of dilute randomly oriented assemblies of the same nanoparticles. The hysteresis loop for $D = 70$ nm is close to the corresponding SW loop, although it has a lower coercive force due to the influence of the curling mode. On the other hand, the coercive force of the vortex particle with $D = 90$ nm is close to zero. Nevertheless, as Fig. 10b shows, the remanent magnetization of dilute randomly oriented assemblies of particles of this type in the range of diameters studied practically coincides with that for the corresponding SW hysteresis loop.

**$a/b = 2.0$**

Calculations of hysteresis loops for the case $a/b = 2.0$ and 'z' type anisotropy were carried out for particles with transverse diameters $D = 80$, 96, and 110 nm. For particles of this type the single-domain diameter was determined [15] as $D_c = 106.0$ nm. Since the upper diameter for the uniform rotation mode for these particles is $D_{cr} = 55.8$ nm, the particles with diameters $D = 80$ and 96 nm are single-domain, but fall in the range of existence of the curling mode, while vortex_p is the ground state of the particle with $D = 110$ nm in zero magnetic field. In addition, for particles of this type in the diameter range $D_c < D \leq 110$ nm, there is also a homogeneous metastable state magnetized along Z axis.

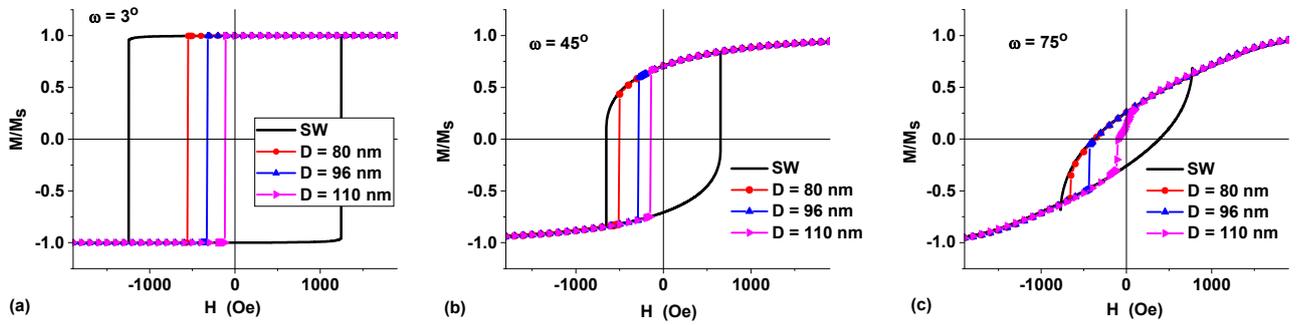

Fig. 11. Comparison of hysteresis loops of particles with 'z' type anisotropy, $a/b = 2.0$ and transverse diameters $D = 80$, 96 and 110 nm at the angles $\omega = 3°$ (a) $\omega = 45°$ (b) and $\omega = 75°$ (c), respectively. SW hysteresis loops are shown as solid curves.

The magnetization reversal process for single-domain nanoparticles with diameters $D = 80$ and 96 nm is similar to that discussed above for single-domain particles with $a/b = 1.5$. In particular, a particle with $D = 80$ nm remains uniformly magnetized for all directions of applied magnetic field. It changes magnetization abruptly due to nucleation of the curling mode, similar to the process shown in Fig. 8a. The hysteresis loops of the particle with $D = 96$ nm show the same behavior at the angles $\omega \leq 82°$. Nucleation of the curling mode in these particles leads to a noticeable decrease in the switching fields of these particles at small and moderate values of $\omega$, as can be seen in Figs. 11a and 11b, respectively. However, for $\omega = 87°$ in the particle with $D = 96$ nm the vortex exists in the field intervals of 400 Oe $\leq H \leq$ 1200 Oe and -1500 Oe $\leq H \leq$ -600 Oe, respectively, whereas the uniform magnetization is restored in this particle in the interval -600 Oe $\leq H \leq$ 400 Oe. Thus, the magnetization reversal in this particle at large $\omega$ is similar to the previously studied case shown in Fig. 8b.

Although the particle with diameter $D = 110$ nm is not single-domain, at the angles $\omega \leq 60°$ it remains uniformly magnetized in applied magnetic field. Magnetization reversal in this particle occurs in small negative magnetic fields in the same way as shown in Fig. 8a. As a result, at $\omega \leq 60°$ the hysteresis loops of this particle differ from the loops for single-domain particles with $D = 80$ and 96 nm only in the smaller value of the switching field (see Figs. 11a and 11b). However, as curve 2 in Fig. 12a shows, at $\omega = 75°$ a vortex appears in this particle already in a large positive magnetic field $H = 1300$ Oe. With a decrease in the field, this vortex evolves, having a large z component of magnetization. In the field range from 50 Oe to -100 Oe, the vortex turns perpendicular to the Z axis and is strongly compressed, so that at $H = -100$ Oe its total magnetization drops to $M_t = 0.1$. But already in the field $H = -120$ Oe, the $M_z$ component changes sign, and its absolute value increases abruptly. Subsequently, the vortex is slowly displaced from the particle and disappears at $H$ = - 1700 Oe. The magnetization reversal goes in this particle in similar way for the angles $\omega$ > 75°.

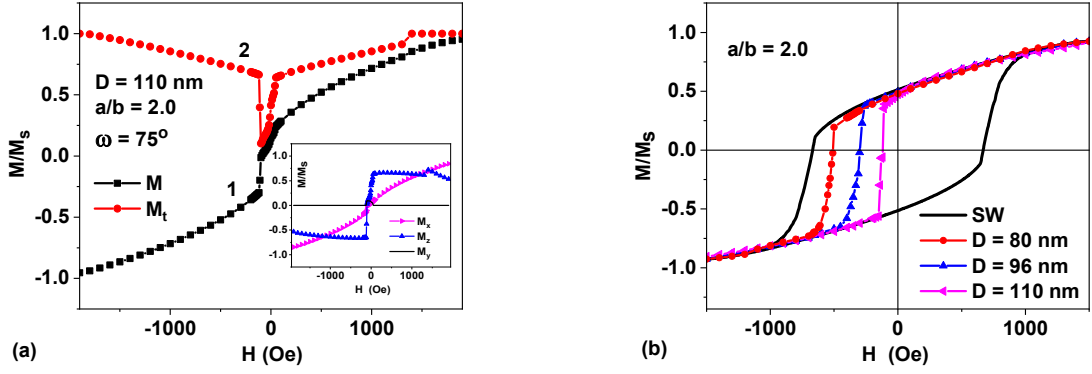

Fig. 12. (a) The magnetization reversal process of vortex particle with diameter $D$ = 110 nm at $\omega$ = 75°. (b) Hysteresis loops of dilute randomly oriented assemblies with aspect ratio $a/b$ = 2.0 and 'z' type anisotropy for particles with transverse diameters $D$ = 80, 96 and 110 nm in comparison with the SW hysteresis loop.

The angular dependence of the switching fields of nanoparticles with aspect ratio $a/b$ = 2.0 is similar to that shown in Fig. 10a for particles with aspect ratio $a/b$ = 1.5. Fig. 12b shows the hysteresis loops of dilute randomly oriented assemblies with aspect ratio $a/b$ = 2.0 and 'z' type anisotropy for particles with diameters $D$ = 80, 96 and 110 nm. Because of the influence of the curling mode the assemblies of single-domain particles with diameters $D$ = 80 and 96 nm have a reduced coercivity compared to that of randomly oriented assembly of SW particles. The hysteresis loop of randomly oriented assembly of vortex particles with $D$ = 110 nm has an even lower coercivity. At the same time, as Fig. 12b shows, the remanent magnetization of dilute randomly oriented assemblies of particles of this type is close to that for the corresponding assembly of SW particles.

It is interesting to compare the results obtained above for elongated spheroidal magnetite particles with hysteresis loops for spherical nanoparticles. The latter were studied in detail [19] for the case when the particle magnetization reversal is carried out by coherent magnetization rotation. A single-domain spherical magnetite particle in the ground state is uniformly magnetized along one of the easy cubic anisotropy axes. The effective anisotropy constant of spherical magnetite particle is equal to $K_{ef} = 2|K_c|/3$, and the critical field of uniform magnetization rotation in the case when the external magnetic field is applied along the easy anisotropy axis is given by Eq. (3). The nucleation field of the curling mode for this particle is given by [17,18]

$$H_n = -\frac{4|K_c|}{3M_s} - C\frac{\gamma_{11}^2}{R^2 M_s} + NM_s, \qquad (10)$$

where $N = 4\pi/3$, $R$ is the sphere radius, $\gamma_{11}$ = 2.0816 is the smallest root of the derivative of the spherical Bessel function. Equating formulas (3) and (10), one finds the critical diameter, $D_{cr} = 2\gamma_{11}L_{ex}/\sqrt{N}$, above which the curling mode becomes the easiest one for a spherical particle at $\omega$ = 0. For spherical magnetite nanoparticle one obtains $D_{cr}$ = 63.9 nm, while the single-domain diameter of this particle was determined [15] as $D_c$ = 70.4 nm.

As Fig. 13a shows, the hysteresis loops of randomly oriented assemblies of vortex magnetite particles with diameters $D$ = 75 and 85 nm have a reduced coercivity compared to that of the corresponding assembly of SW particles. In contrast to assemblies of spheroidal particles, the remanent magnetization of assemblies of spherical nanoparticles decreases significantly with increasing particle diameter. Note that for spherical nanoparticles it is necessary to take into account the weak dependence of quasi-static hysteresis loops on the azimuthal angle $\psi$. The hysteresis loops of randomly oriented assemblies shown in Fig. 13a are averaged over this angle.

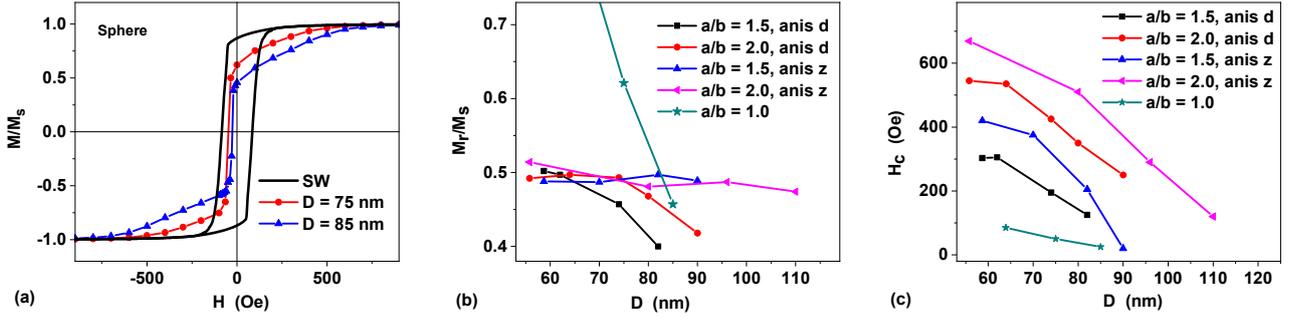

Fig. 13. (a) Hysteresis loops of dilute randomly oriented assemblies of spherical, non single-domain magnetite nanoparticles with diameters $D$ = 75 and 85 nm in comparison with the hysteresis loop of randomly oriented assembly of single-domain nanoparticles [19]. Comparison of remanent magnetization (b) and coercivity (c) of dilute randomly oriented assemblies with different combined anisotropy and particle aspect ratio.

In Fig. 13b we compare the remanent magnetization of the dilute randomly oriented assemblies with various aspect ratios and types of anisotropy, studied in this work. In the diameter range studied, the remanent magnetization of the assemblies with 'z' type anisotropy is close to the value $M_r/M_s$ = 0.5, characteristic of assemblies with a purely uniaxial type of magnetic anisotropy. For assemblies of particles with 'd' type anisotropy, the remanent magnetization noticeably decreases with increasing transverse particle diameter. At the same time, the remanent magnetization of assemblies of spherical nanoparticles drops sharply from the known value $M_r/M_s$ = 0.866 [25] for randomly oriented assembly of single-domain spherical magnetite particles to the value $M_r/M_s$ = 0.46 for the assembly of vortex nanoparticles with diameter $D$ = 85 nm.

The dependence of the coercive force of the same assemblies on the transverse particle diameter is shown in Fig. 13c. The coercivity of assemblies of particles with aspect ratio $a/b$ = 2.0 is noticeably higher than that for particles with $a/b$ = 1.5, which is naturally explained by the influence of the particle shape anisotropy energy. In addition, at the same aspect ratio, the coercive force of assemblies with anisotropy of the 'z' type is higher than that of assemblies with anisotropy of 'd' type. The assemblies of spherical nanoparticles have the lowest coercive force. In all cases, the coercive force of the assemblies decreases noticeably with increasing transverse particle diameter.

## 4. Conclusions

The study of the quasi-static hysteresis loops of an assembly of magnetic nanoparticles provides significant information about the magnetic properties of the assembly. Unfortunately, quasi-static hysteresis loops of assemblies with diameters on the order of or slightly larger than the single-domain diameter $D_c$ have still not been investigated in detail. Meanwhile, the properties of assemblies of iron oxides particles being in a vortex micromagnetic state have recently attracted interest for use in biomedicine [1-5]. In addition, submicron nanoparticles of magnetic iron oxides are one of the main objects of research in paleomagnetism [6-14].

In this work we study quasi-static hysteresis loops of spherical and spheroidal magnetite nanoparticles with aspect ratios $a/b$ = 1.5 and 2.0 and different types of combined magnetic anisotropy using numerical simulation. The hysteresis loops of dilute oriented assemblies are calculated depending on the spherical angle $\omega$ the magnetic field makes with the particle symmetry axis. The hysteresis loops of the corresponding randomly oriented assemblies are obtained by averaging over this angle. For particles of each type, we estimate critical diameter $D_{cr}$, above which the magnetization curling [17, 18] is the easiest mode of particle magnetization reversal. Calculations of hysteresis loops of assemblies of each type are carried out both for single-domain nanoparticles in the diameter range $D_{cr} < D \leq D_c$, and for vortex particles with diameters $D > D_c$. The results obtained are compared with the hysteresis loops of Stoner–Wohlfarth type particles, where the magnetization reversal is carried out by the coherent rotation of magnetization, and the quasi-static hysteresis loops do not depend on the particle diameter.

For assemblies of spheroidal magnetite particles with aspect ratio $a/b \geq 1.5$ the influence of uniaxial magnetic anisotropy arising due to the particle shape anisotropy energy is found to be significant. As a result, the quasi-static hysteresis loops of such assemblies show only a weak dependence on the azimuthal angle $\psi$. The predominance of uniaxial magnetic anisotropy is especially characteristic of particles with combined anisotropy of the 'z' type. In particular, they have single-domain diameters significantly exceeding the $D_c$ value for spherical particle.

In accordance with the analytical results [17,18], it is shown that in the diameter range $D_{cr} < D \leq D_c$ for small angles, $\omega \leq 30°$, the inhomogeneous curling mode nucleates much earlier than the uniform rotation mode. In addition, for particles with combined anisotropy of type 'd', the nucleation of the curling mode is not accompanied by immediate magnetization reversal of the particle, as can be seen in Fig. 1a ($D$ = 62 nm) and Fig. 4a ($D$ = 64, 72 nm), respectively. For these particles the vortex is stabilized in a certain intervals of field so that the switching field of these particles $|H_{sw}| > |H_{curl}|$. At the same time, for particles with combined anisotropy of the 'z' type, stabilization of the vortex is not observed in the diameter range studied (see Fig. 7a, $D$ = 70, 82 nm and Fig. 11a, $D$ = 80, 96 nm). It is worth mentioning the appearance of a vortex in the intermediate ranges of magnetic fields for single-domain particles with combined anisotropy of the 'z' type. As Fig. 8b shows, it happens for particles with $D \sim D_c$ at sufficiently large angles $\omega$.

It is worth noting the significant difference of the magnetization reversal processes in vortex particles for small and moderate angles, $\omega \leq 45 - 60°$, and for large ones, $\omega \geq 75°$, respectively. For vortex particles with 'd' type anisotropy, the vortex appears in a sufficiently large positive magnetic field $H_{nv}$, which leads to reduced values of the remanent magnetization of oriented assemblies of such particles for angles $\omega \leq 45°$. However, for vortex particles with 'z' type anisotropy and aspect ratio $a/b$ = 1.5 for angles $\omega \leq 60°$ the transverse vortex appears only in small intervals of negative fields. For these particles with $a/b$ = 2.0 for angles $\omega \leq 60°$ magnetization reversal occurs in a single jump, similar to the behavior of single-domain particles with 'z' type anisotropy. However, for large angles, $\omega \geq 75°$, vortex exists in particles of this type in a wide range of magnetic fields $|H| \leq H_{nv}$. In addition, the magnetization reversal processes of spheroidal magnetite nanoparticles depend on the presence of metastable micromagnetic states in such particles in low magnetic fields. A common property of assemblies of vortex particles is a low coercive force, which decreases with increasing transverse particle diameter. At the same time, the remanent magnetization of randomly oriented assemblies of vortex particles in the studied cases remains close to that for randomly oriented assemblies of SW particles.

The theoretical calculations of quasi-static hysteresis loops of dilute assemblies of vortex and single-domain magnetite particles with diameters close to the single-domain diameters demonstrate interesting features of the magnetization reversal processes in such particles in applied magnetic field. In our opinion, these results deserve experimental confirmation.

**References**

1. D. Cao, H. Li, L. Pan, J. Li, X. Wang, P. Jing, X. Cheng, W. Wang, J. Wang, Q. Liu, Sci. Reports 6 (2016) 1, https://doi.org/10.1038/srep32360
2. N.A. Usov, M.S. Nesmeyanov, V.P. Tarasov, Sci. Reports 8 (2018) 1224, https://doi.org/10.1038/s41598-017-18162-8
3. H. Gao, T. Zhang, Y. Zhang, Y. Chen, B. Liu, J. Wu, X. Liu, Y. Li, M. Peng, Y. Zhang, G. Xie, F. Zhao, H.M. Fan, Journal of Materials Chemistry B, 8 (2020) 515, https://doi.org/10.1039/C9TB00998A
4. G.R. Lewis, J.C. Loudon, R. Tovey, Y.H. Chen, A.P. Roberts, R.J. Harrison, P.A. Midgley, E. Ringe, Nano Letters 20 (2020) 7405, https://dx.doi.org/10.1021/acs.nanolett.0c02795
5. V.C. Karade, A. Sharma, R.P. Dhavale, S.R. Shingte, P.S. Patil, J.H. Kim, D.R. Zahn, A.D. Chougale, G. Salvan, P.B. Patil, Sci. Reports 11 (2021) 1, https://doi.org/10.1038/s41598-021-84770-0
6. D. J. Dunlop, O. Ozdemir, Rock magnetism, fundamentals, and frontiers, Cambridge Univ. Press, Cambridge, U.K., 1997.

7. R.J. Harrison, R.E. Dunin-Borkowski, A. Putnis, Proc. Natl. Acad. Sci. USA 99 (2002) 16556, https://doi.org/10.1073_pnas.262514499.
8. A. R. Muxworthy, W. Williams, Journal of Geophysical Research, 111 (2006) B12S12. http://dx.doi.org/10.1029/2006jb004588
9. I. Lascu, J.F. Einsle, M.R. Ball, R.J. Harrison, Journal of Geophysical Research: Solid Earth, 123 (2018) 7285. https://doi.org/10.1029/2018JB015909
10. L. Nagy, W. Williams, L. Tauxe, A.R. Muxworthy, Geochemistry, Geophysics, Geosystems, 20 (2019) 2907. http://dx.doi.org/10.1029/2019gc008319
11. T. P. Almeida, T. Kasama, A. R. Muxworthy, W. Williams, L. Nagy, R. E. Dunin-Borkowski, Geophys. Res. Lett. 41 (2014) 7041, https://doi.org/10.1002/2014GL061432.
12. L. Nagy, W. Williams, A. R. Muxworthy, K. Fabian, T. P. Almeida, P. O. Conbhui, V. P. Shcherbakov, Proc. Natl. Acad. Sci. USA 114 (2017) 10356, https://doi.org/10.1073/pnas.1708344114.
13. K. Fabian, V.P. Shcherbakov, Geophysical Journal International, 215 (2018) 314, https://doi.org/10.1093/gji/ggy285.
14. T.P. Almeida, A.R. Muxworthy, A. Kovács, W. Williams, L. Nagy, P. Ó Conbhuí, C. Frandsen, R. Supakulopas, R.E. Dunin-Borkowski, Geophys. Res. Lett. 43 (2016) 8426, https://doi.org/10.1002/2016GL070074.
15. N.A. Usov, O.N. Serebryakova, J. Magn. Magn. Mater. 588 (2023) 171345, https://doi.org/10.1016/j.jmmm.2023.171345.
16. E.C. Stoner, E.P. Wohlfarth, Phil. Trans. R. Soc. London, Ser. A240 (1948) 599.
17. W.F. Brown, Jr., Micromagnetics, Wiley-Interscience, New York - London, 1963.
18. A. Aharoni, Introduction to the Theory of Ferromagnetism, Clarendon Press, Oxford, 1996.
19. N.A. Usov, S.E. Peschany. J. Magn. Magn. Mater. 174 (1997) 247.
20. E. Schabes, J. Magn. Magn. Mater. 95 (1991) 249, https://doi.org/10.1016/0304-8853(91)90225-Y.
21. J. Fidler, T. Schrefl, J. Phys. D: Appl. Phys. 33 (2000) R135, https://doi.org/10.1088/0022-3727/33/15/201.
22. A. Aharoni, J. Appl. Phys. 63 (1988) 5879, http://dx.doi.org/10.1063/1.340280.
23. A. Aharoni, IEEE Trans. Magn. 25 (1989) 3470.
24. M. Abramowitz, I.A. Stegun, Handbook of Mathematical Functions, National Bureau of Standards, 1964.
25. R Gans, Ann. Phys. 15 (1932) 28.